# Cross Script Hindi English NER Corpus from Wikipedia

Mohd Zeeshan Ansari, Tanvir Ahmad and Md Arshad Ali

Department of Computer Engineering, Jamia Millia Islamia, New Delhi, India.

mzansari@jmi.ac.in, tahmad2@jmi.ac.in, arshad1020@gmail.com

**Abstract.** The text generated on social media platforms is essentially a mixed lingual text. The mixing of language in any form produces considerable amount of difficulty in language processing systems. Moreover, the advancements in language processing research depends upon the availability of standard corpora. The development of mixed lingual Indian Named Entity Recognition (NER) systems are facing obstacles due to unavailability of the standard evaluation corpora. Such corpora may be of mixed lingual nature in which text is written using multiple languages predominantly using a single script only. The motivation of our work is to emphasize the automatic generation such kind of corpora in order to encourage mixed lingual Indian NER. The paper presents the preparation of a Cross Script Hindi-English Corpora from Wikipedia category pages. The corpora is successfully annotated using standard CoNLL-2003 categories of PER, LOC, ORG, and MISC. Its evaluation is carried out on a variety of machine learning algorithms and favorable results are achieved.

## 1    Introduction

Named Entity Recognition (NER) is a significant task in Information Extraction which identifies and classifies information elements called Named Entities (NEs) in a given sample of text. The term Named Entity was coined at the 6th Message Understanding Conference (MUC) to encourage the importance of semantic identification of persons, locations and organizations in natural language text [1]. NER is far from being a solved task due to large dissimilarity around its concept which shows impact on its applications and evaluations [2,7,9]. The efforts in the progress of NER systems for Indian languages, primarily Hindi faced adversities due to unavailability web resources, as most of them are in English. Moreover, language mixing is common phenomenon on social media which increases considerable amount of difficulty in its corpus creation. The data analysis of Indian social networking sites, blogs and chatting applications show great deal of anglicism in its text [3]. The key challenge in the construction and evaluation of Indian NER systems is the unavailability of standard evaluation corpora for mixed lingual text. As an example with binary language set, a mixed bilingual text is one that contains Hindi, English and Romanized Hindi. In motivation to boost the advancements in research and studies on Indian mixed lingual information extraction, we attempt to build a cross script Hindi-English NER corpus from Wikipedia. This work will enable to develop informatics, probably to build systems for extraction of named entities from Indian social media content. It will empower the cyber media regulatory authorities to tackle the severe problem of using social media platform for intimidation. Moreover, it may be helpful in detection of fake and offensive messages spread against a person or an organization. To the best of our knowledge, our work is the first attempt to use Wikipedia for building such corpora.

To build the cross script Hindi English NER corpora (Hi-En-WP), we explored Wikipedia and subsequently identified specific category pages having substantial amount of links to the Wikipedia pages pertaining to Hindi belt of India [17]. We considered the entities targeted in such pages as Hindi-English cross script due to its origin from Hindi linguistic sections, but written in Roman script. The automatically extracted collection of multi-token expressions from the links of those Wikipedia content pages are considered as probable NEs. Each of these probable NEs is assigned a confidence score based on its attributes. The probable NEs with certain level of confidence score are selected for inclusion in the corpora. Further, the constructed corpora is evaluated on various machine learning algorithms.

## 2    NER and Wikipedia

Wikipedia is an open and collaborative multilingual encyclopedia contributed by several collaborators currently having 5.6 million English articles [21]. Constantly, the articles are updated and new articles are added by its collaborators. About 74% of Wikipedia articles fall under the category of named entity classes [4], therefore, we consider Wikipedia links among articles to construct the Hi-En-WP NER annotated corpus. Wikipedia includes content pages which contain concepts and facts about the article, category pages provides a list of content pages into several classes based on specific criteria and disambiguation pages help to locate different content pages with same title.

Wikipedia is an abundant resource for generation of different types of multilingual, cross lingual, cross script and language independent corpus, etc., its markup has been used extensively to automatically generate NER annotated corpus for training machine learning models [4-6,11-14,19]. The latest involvement using Wikipedia is the portable cross lingual NER for low resource languages using translation of an annotated NER corpus from English [12,19]. Another approach to cross lingual and language independent corpora is to learn a model on language independent features of a source language and test the model on other languages using same features [13]. Nothman, et al. (2008) [5] constructed a massive English NER corpus by the classification of Wikipedia articles to its category types by mapping them to CoNLL-2003 NER tagset. A similar approach to massive multilingual NER corpus is found in [14]. A hybrid approach to generate parallel sentences with NE notations reveal strong results on Wikipedia dataset [19]. Kazama and Torisawa (2007) [15] use the external knowledge along with the analysis of first sentence in Wikipedia articles to infer the entity category. Cross script NER corpus from social media in Bengali written in Roman script is also evaluated on various models [10].

Domain specific gazetteers for Indian languages are developed using transliteration and context pattern extraction. Saha et al. (2008) [8] proposed an approach to develop NE corpora for Indian languages from various web resources using transliteration and context pattern extraction. Sequence labelling was employed on English-Hindi and English-Tamil to obtain suitable results. A corpora for Bengali-English was derived from social media and was evaluated using contextual features on Tourism and Sports domain. Our approach to classify the articles is based on computation of the entity confidence score for each multi-token extracted as a probable NE and tag the probable NE having certain high confidence score to CoNLL-2003 [21] NE classes. Eventually, there is no standard cross script Hindi-English NER corpus available, we are first to utilize Wikipedia for such corpus preparation. Except for Bengali-English Code Mixed Cross Script Named Entity corpora which is extracted from social media content [10], to the best of our knowledge, no Hindi-English cross script automatically built corpora is available which is extracted from Wikipedia.

## 3 Hindi English Wikipedia Corpus Generation

The automatic generation of Hi-En-WP corpora involves three tasks to be performed: extracting entities from a webpage, generating a confidence score for each entity for selection into named entity set, and finally annotating them with the appropriate tag. To achieve this, we parsed 13 Wikipedia category pages and collected 7285 hyperlinks. The category pages that we selected to initiate the extraction process were those class of Wikipedia pages that contain surplus information that is closely related to NEs and positively about Hindi speaking regions of India [17]. We strongly take into account the assumption that the entities present on these pages are prominently Hindi NEs. This assumption is based on human assessment that the information on such pages is based on Indian background especially from Hindi linguistic majority sections of India.

### 3.1 Corpus Acquisition

Wikipedia being a huge source of information, its articles comprise of: topic and its comprehensive summary in paragraphs and images; reference to reliable resources; and hyperlinks, also called wikilinks to other articles. Our method takes the advantage of wikilinks between the articles from which linktext is extracted. Since wikilinks are links to articles, it may be considered as a named entity. This approach is similar to Nothman et al (2008) [4] to generate the NER data from wikilinks. A total number of 7285 tokens and multi-tokens expressions were extracted from the links by parsing the 13 identified Wikipedia webpages. We consider these expressions as probable NEs after the removal of duplicates from the set of tokens obtained.

In order to distinguish among the tokens an intuitive approach is to calculate the distribution of different type of tokens. To make this achievable, we grouped them according to

| **Table 1.** Hi-En-WP Corpus statistics. | |
|---|---|
| | Hi-En-WP |
| Wikipedia category pages accessed | 13 |
| Hyperlink based tokens, multi-tokens generated | 7285 |
| Probable NEs after duplicate removal | 5401 |
| Total selected NEs | 2916 |
| NE density | 40.02 % |

| **Table 2.** Distribution of NER tagset of Hi-En-WP. | | |
|---|---|---|
| NER tagset | % | Number of entities |
| PER | 65 | 1883 |
| LOC | 17 | 492 |
| ORG | 13 | 388 |
| MISC | 5 | 153 |
| Overall | 100 | 2916 |

**POS tags.** We consider the supposition that NEs are often NNP, moreover, the common nouns, verbs and rest of the POS tags generated help to filter out the non NEs.

**Wordtypes.** We adopt a subset of *wordtypes* proposed by Collins (2002) [16], which maps uppercase letters to A, lowercase to a, and, digits to 0.

The brief statics of the generated HI-En-WP corpus is presented in Table 1.

### 3.2 Confidence score

According to the POS tags and wordtypes assigned to each expression, the confidence score is calculated for the selection of the NEs to be annotated. Against each probable NE two scores are generated: (1) POS tag score is assigned a value of 1 if the tag is NNP, NNS or NN, and 0 for rest of the tags. (2) wordtypes score is assigned a value of 1 if the type starts with A, or contains all A, and 0 for rest of the cases. In this way, a confidence score is constituted by the sum of both the scores for all the probable NEs. A positive confidence score achieved by a probable NE is marked selected for the corpus. The probable NEs which are allocated a zero confidence score as discarded from selection. In this way, we obtained 2916 entities ready for annotation into NER tagset. More complex score can be taken into account by the different attributes relevant for named entity classification. Also, the usage of different cases of POS tags, wordtypes can be explored. The task is easily scalable to generate massive corpus by the consideration of more number of source Wikipedia pages.

### 3.3 NER tagsets

Two different annotators manually tagged almost 2916 NEs to course grained labels based on CoNLL-2003 tagset, i.e., PER, LOC, ORG and MISC [21]. Both the annotators were proficient in Hindi as well as in English. The accurately labelled corpus achieved the inter annotator agreement over 98%. A little disagreement was over MISC tags which were mutually resolved. The predominantly occupied class over the whole set is PER which is 65%, LOC and ORG tagged NEs are almost distributed similarly, MISC is only 5 percent, see Table 2. We experimented with this course-grained corpus in various configurations of train and test set over different machine learning approaches

**Table 3.** Precision, recall and F-score for each classifier on evaluation of Hi-EN-WP corpora.

| Evaluation Algorithm | P | R | F |
|---|---|---|---|
| Logistic Regression | **0.89** | **0.89** | **0.88** |
| SVM | 0.88 | 0.88 | 0.88 |
| Random Forest | 0.85 | 0.86 | 0.84 |
| Naïve Bayes | 0.83 | 0.82 | 0.79 |
| SGD Classifier | 0.81 | 0.82 | 0.81 |

## 4      Evaluation of Hi-En-WP Corpus

Machine Leaning algorithms are quite successful for NER training and prediction. The algorithms make use of patterns, contextual information, orthographic and linguistic features in order to train the models. The annotated training corpora common towards NER task is Newswire text, CoNLL-2003 shared task data, the MUC7 dataset, BBN, etc. We evaluated the performance of word level Hi-En-WP corpus on general classification algorithms using the collection of all character n-grams for n=1-5 as feature set. Amongst Logistic Regression, SVM, Naïve Bayes, Random Forest and Stochastic Gradient Classifier, Logistic Regression seemed most effective, in all the cases, as shown in Table 3.

   The analysis of class wise precision, recall and F-score on Logistic Regression classifier is shown in Table 4. When trained with MISC as a separate class, high values for PER suggests that they comparatively easy to identify. It may be due to sufficiently large amount of training data as compared to rest of the tags. Whereas, low precision value for LOC tag suggests that other entities are wrongly classified as location. The MISC F-score is expectedly low, in agreement with the results of Nothman et al (2008) [4]. The variation reflected in F-score among all may be the effect of diversity in linguistic attributes. An increase in accuracy from 89% to 92% is observed when the model is trained without MISC tag which reflects that the confusion is created in data by the inclusion of training examples that belong to MISC tag. The Fig.1 illustrates the effect of varying size of the Hi-En-WP training data. The improving trend of accuracy score on increasing training data sufficiently produces scope to scale the size of corpus in future.

**Table 4.** Class distribution of NEs and average results of five-fold cross validation on logistic Regression. Overall results are micro-averaged.

| Class | % | With MISC | | | Without MISC | | |
|-------|---|---|---|---|---|---|---|
| | | P | R | F | P | R | F |
| PER | 65 | 0.93 | 0.97 | 0.95 | 0.95 | 0.97 | 0.96 |
| LOC | 17 | 0.80 | 0.89 | 0.84 | 0.86 | 0.89 | 0.87 |
| ORG | 13 | 0.82 | 0.77 | 0.79 | 0.91 | 0.77 | 0.84 |
| MISC | 5 | 0.74 | 0.28 | 0.41 | - | - | - |
| Overall | | 0.89 | 0.89 | 0.88 | 0.93 | 0.93 | 0.92 |

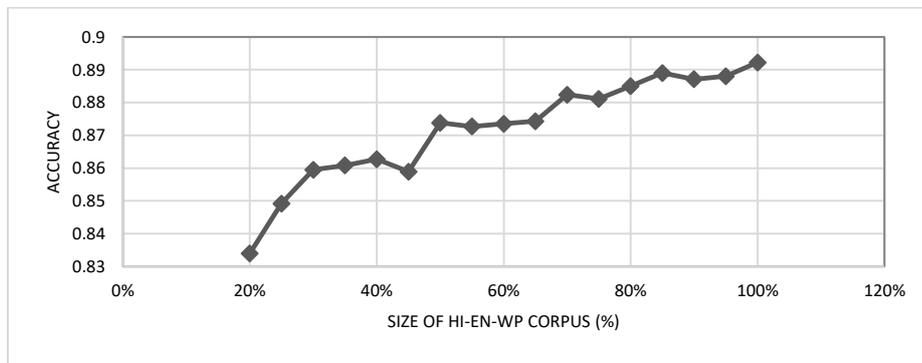

**Fig. 1.** Effect of varying Hi-En-WP corpus size on accuracy.

# 5 Conclusion and Discussion

We prepared a cross script Hindi English NER annotated corpora automatically from Wikipedia for mixed lingual NER tasks. We exploit the wikilinks of Indian context Wikipedia pages to extract the entities. The manually annotated corpora is examined against POS tags and Wordtypes in order to remove non-named entities. We retrieve fair number of named entities which is largely occupied by persons. We evaluated the corpus against itself across several machine learning algorithms and obtained promising results. There is much scope for improvement of the corpus, firstly, more number of attributes of wikilinks can be explored to formulate a strong confidence score. Secondly, the corpus can be distributed according to the domain to which each named entity belongs. Finally, the cross corpus evaluation of Hi-En-WP corpus can be carried out against CoNLL-2003, MUC-7 and other named entity corpus in order to perform a comparative analysis.